\documentstyle[twoside,fleqn,espcrc2,epsf]{article}

\newcommand{\AmS}{{\protect\the\textfont2
  A\kern-.1667em\lower.5ex\hbox{M}\kern-.125emS}}

\newcommand{\bee}{\begin{equation}}
\newcommand{\ee}{\end{equation}}
\newcommand{\beea}{\begin{eqnarray}}
\newcommand{\eea}{\end{eqnarray}}

\newcommand{\PL}[3]{{Phys. Lett.} {\bf #1} {(19#2)} #3}
\newcommand{\PR}[3]{{Phys. Rev.} {\bf #1} {(19#2)}  #3}
\newcommand{\NP}[3]{{Nucl. Phys.} {\bf #1} {(19#2)} #3}

\title{Exceptional Configurations with the Clover Action}

\author{Thomas DeGrand, Anna Hasenfratz and
        Tam\'as G. Kov\'acs \\[2mm]
        Department of Physics, 
        University of Colorado\\ Boulder, CO 80309-390, USA}

\begin{document}

\begin{abstract}
We study exceptional modes of both the Wilson and the clover
action in order to understand why quenched
clover spectroscopy suffers so severely from exceptional
configurations. We show that a large clover coefficient 
can make the exceptional modes extremely localized and
thus very sensitive to short distance fluctuations.
We contrast this with the case of the Wilson action
where exceptional modes correspond to large instantons.
These modes are broadly extended and suffer much less
from discretization errors. 
\end{abstract}

\maketitle

\section{Introduction}

Exceptional configurations have recently been shown to be due
to real eigenvalues of the Dirac operator occuring close to 
minus the bare quark masses used in spectroscopy
\cite{Fermilab1,Fermilab2}. These real modes are the lattice
counterparts of the continuum zero modes, shifted away from
zero due to the additive mass renormalization.
A chiral improvement of the lattice Dirac operator is 
expected to decrease the shift of zero-modes as well as their spread.
It is thus surprising that this 
does not seem to be true for the simplest improvement on the
Wilson action, the clover action. Although the clover term 
reduces the additive mass renormalization as compared to the
Wilson action, the problem with exceptional configurations 
appears to be more severe for the clover than for the Wilson
action; at least with the non-perturbatively determined value
of the clover coefficient $c_{sw}$ \cite{Gockeler}. 

In the present paper we show why this happens by studying
how the real modes of the Wilson Dirac operator change when the
clover term is gradually turned on. We also make a first step towards
establishing a connection between exceptional modes and instantons
both for the Wilson and the clover case.

\section{Wilson Eigenmodes and Instantons}
   \label{sec:inst}

Contrary to what happens in the continuum,
the lattice Wilson Dirac operator does not
have an exact zero-mode in the presence of an instanton.
Instead, it has a definite chirality
real eigenmode appearing in the physical branch of the 
spectrum. On smooth one-instanton configurations the location
of the real eigenvalue depends on the size of the corresponding
instanton. Large instantons have modes closer to zero, small
ones have modes farther away from zero, in the direction of 
the real doubler modes\cite{Nar_smooth}. 

\begin{figure}[h!tb]
\begin{center}
\leavevmode
\epsfxsize=70mm
\epsfbox{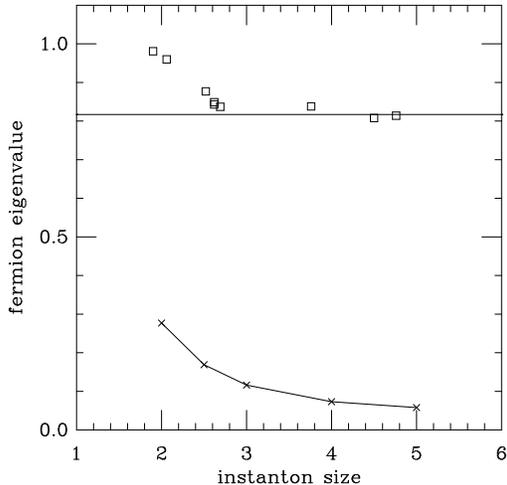}
\end{center}
\caption{The instanton size (in units of the lattice spacing)
vs.\ the eigenvalue of the
corresponding fermionic real mode on smooth instantons
(crosses) and on real Monte Carlo generated configurations
at $\beta=6.0$ (squares). The horizontal line indicates 
$-m_c$ for the $\beta=6.0$ quenched ensemble.}
\label{fig:rho_vs_pole}
\end{figure}

A similar connection between the instanton size and the location
of the corresponding fermionic real mode can be found also on
Monte Carlo generated gauge configurations. In Fig.\ 
\ref{fig:rho_vs_pole} we show the instanton
size versus the corresponding fermionic eigenvalue both
for smooth instantons and instantons identified on real Monte Carlo
generated configurations at $\beta=6.0$ on $12^4$ lattices.
The instanton sizes were measured using the method of Ref.\ 
\cite{SU3INST} and the real modes were extracted exactly as
in Ref.\ \cite{DHK_ape}. To find the corresponding instanton
for each real mode, we compared the quark density of the given
mode to the profile of the charge density. 
Fig.\ \ref{fig:rho_vs_pole} contains results 
only from uniquely identifiable modes.

\section{Eigenmodes with the Clover Action}
    \label{sec:clover}

\subsection{Smooth instantons}

We have seen that on fine enough lattices with the Wilson 
action there is a strong correlation between the 
instanton size and the eigenvalue of the corresponding fermionic
real mode; (near) exceptional modes correspond to large
instantons. We shall now explore what happens in the presence
of the clover term. We start by looking at smooth single instanton
configurations. 

\begin{figure}[h!tb]
\begin{center}
\leavevmode
\epsfxsize=70mm
\epsfbox{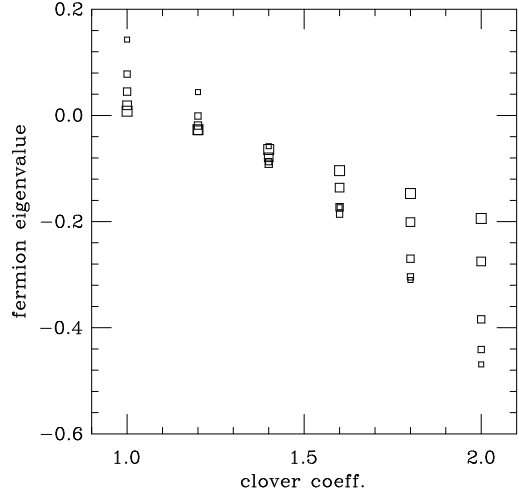}
\end{center}
\caption{The real fermionic eigenvalue versus the clover coefficient
on instantons of sizes $\rho/a=1.2, 1.4, 1.6, 2.0, 2.5$. Bigger symbols
correspond to larger instantons.}
\label{fig:clover}
\end{figure}

In Fig.\ \ref{fig:clover} we plot how the fermion eigenvalues
associated to instantons of various sizes
change as a function of the clover coefficient. For better 
legibility in the figure we include only the range of 
$c_{sw} \geq 1.0$. For the Wilson action $(c_{sw}=0)$ the 
eigenvalues corresponding to instantons of size $1.2 \leq \rho
\leq 2.5$ are spread between 0.19 and 0.86.

As $c_{sw}$ increase from zero to the tree level value, $c_{sw}=1$, 
all the modes move closer 
to zero and their spread decreases. If $c_{sw}$ is further increased,
the modes start to pass through zero and also the trajectories 
corresponding to instantons of different sizes start to cross 
one another. By $c_{sw}=1.4$ the order of all the eigenvalues 
for instanton sizes $\geq 1.5a$ has been reversed. As $c_{sw}$
is increased, modes of smaller and smaller instantons cross over
to the other side of the distribution.

\subsection{Monte Carlo Configurations}

We studied an ensemble of $\beta=5.7$ $6^3\times16$ lattices.
We followed how the wave function of a few typical very exceptional
modes changed with the clover coefficient starting from the
Wilson action $(c_{sw}=0.0)$ up to the non-perturbatively determined
value $(c_{sw}=2.25)$. 
 
With the Wilson action we found that the quark density in
modes close to $-m_c$ has an extended broad peak and for modes 
towards the doublers the peaks get narrower and sharper. This is
in complete agreement with \cite{Nar_size}. 

Let us now follow what happens to the wave function of a typical
exceptional mode as the clover term is gradually turned on.
Up to $c_{sw}=1.0$ the main peak of the quark density remains
at the same location. It only slowly gets narrower: at 
$c_{sw}=1.0$ its width is 2.3. Increasing $c_{sw}$ further,
another peak appears and its relative significance 
increases with the clover coefficient. By $c_{sw}=2.25$ 
the wave function is completely concentrated on a very
sharp peak (of width $\approx 1$).
The wave function of this $c_{sw}=2.25$ exceptional
mode is very similar to that of a Wilson mode lying halfway between
the physical and the doubler branch.

We can now easily describe
qualitatively how the exceptional modes change with the clover
coefficient.  The Wilson modes
can be roughly thought of as being concentrated on single instantons.
As the clover term is turned on, 
the zero modes corresponding to different instanton sizes get closer and the 
quark wave function spreads
over several topological objects.
We saw similar behavior on a subset of the
 Fermilab \cite{Fermilab1,Fermilab2}
configurations given to us by H. Thacker.
If $c_{sw}$ is further increased,
the zero modes separate again, but this time 
the well localized modes (corresponding to smaller instantons) 
have smaller eigenvalues. The relative
importance of the broader peaks decreases and the mode can become
entirely concentrated in a very sharp peak. We want to emphasize 
however that on a given ensemble of gauge configurations at a given
value of $c_{sw}$ different (near) exceptional modes can look
qualitatively very different. Some of the modes are entirely
concentrated in a sharp peak --- typically these are the most exceptional
ones --- some have several peaks of various widths. 

The sharply peaked modes look very much like (near) doubler
modes of the Wilson action and 
the corresponding eigenvalues and the way they change 
with the clover term can be very sensitive to the fluctuations
on the shortest distance scale. 
This is the reason why the clover 
coefficient cannot be optimized to minimize the spread of the
real eigenvalues. On the other hand, this can be done if
the shortest scale fluctuations are ``tamed''. This is
the idea behind our recently proposed new fermion action
with fat gauge links and an optimized clover coefficient \cite{DHK_ape}.
The role of the fattening is to make the fermion modes less
sensitive to the short distance fluctuations. We checked the 
wave functions of the most exceptional modes of our fat link
action and in our sample we never encountered any ``anomalously''
sharply localized mode. 


We expect that minimizing the 
spread of real eigenvalues will also improve the situation 
with exceptional configurations. This is indeed what happens. 
On a set of 40 $6^3\times16$ Wilson $\beta=5.7$ configurations
we determined all the exceptional modes both for our optimized 
fat link action and the thin link clover action with $c_{sw}=2.25$.
For the fat link action all the modes occurred at pion masses
$m_{\pi}<0.3$ which corresponds to $m_{\pi}/m_{\rho} \leq 0.38$.
whereas the clover action had five exceptional
modes above $m_{\pi}=0.3$ at $m_{\pi}= 0.42, 0.53, 0.39, 0.30,
0.62$ (in lattice units).

\section{Conclusions}

In the present paper we gave an explanation of why quenched 
clover spectroscopy on coarse lattices suffers so severely 
from exceptional configurations. Our main result is that increasing
the clover coefficient can make exceptional modes extremely  
localized and thus very sensitive to short distance
fluctuations. Therefore our results are quite 
worrisome for simulations done on coarse lattices with the clover
action. It would be worth repeating this study on a finer lattice
where instantons can be explicitly identified and correlated
with peaks of the eigenmodes. 

This work was supported by the US Department of Energy.


\end{document}